\documentclass[ASNA,twocolumn]{USG}

\usepackage[utf8]{inputenc}
\usepackage[T1]{fontenc}
\usepackage{anyfontsize}
\usepackage{xcolor}
\usepackage{colortbl}
\usepackage{graphicx}
\usepackage{dcolumn}
\usepackage{tikz}
\usepackage{gensymb}
\usepackage{amsmath}
\usepackage{amssymb}
\usepackage{amsthm}
\usepackage{steinmetz}
\usepackage{ragged2e}
\usepackage{ulem}
\usepackage{soul}
\usepackage{pdfpages}
\usepackage{hyperref}

\graphicspath{{./images/}}
\setstcolor{red}
\DeclareUnicodeCharacter{0308}{\"}

\articletype{RESEARCH ARTICLE}
\articledoi{}% required by the USG class footer (DOI is assigned by the journal)

\makeatletter
\IfFileExists{images/ORCID_Logo.pdf}{}{%
  \IfFileExists{images/ORCID_Logo.png}{}{\let\orchid\@gobble}}
\makeatother
\begin{document}

\title{Superconducting Fluctuation Driven Quantum Transport in Bi$_2$Sr$_2$CaCu$_2$O$_{8+\delta}$ Thin Films}

\author[1]{Sohini Guin}[https://orcid.org/0009-0001-1817-3765]
\author[1]{Naresh Shyaga}[https://orcid.org/0009-0001-4433-5393]
\author[1]{Jagadish Rajendran}[https://orcid.org/0000-0003-2575-6202]
\author[1]{Aryaman Das}[https://orcid.org/0009-0003-1228-7061]
\author[1]{Subhransu Kumar Negi}[https://orcid.org/0009-0008-5258-979X]
\author[2]{Saisab Bhowmik}
\author[1]{Pankaj Bhardwaj}[https://orcid.org/0000-0001-8341-525X]
\author[2]{Chandni Usha}
\author[1]{Dhavala Suri}[https://orcid.org/0000-0001-7037-2721]

\address[1]{\orgdiv{Centre for Nanoscience and Engineering, }\orgname{Indian Institute of Science, Bengaluru 560012, }%
\orgaddress{\state{Karnataka, }\country{India}}}

\address[2]{\orgdiv{Instrumentation and Applied Physics, }\orgname{Indian Institute of Science, Bengaluru 560012, }%
\orgaddress{\state{Karnataka, }\country{India}}}

\corres{Dhavala Suri (\email{dsuri@iisc.ac.in})}


\abstract[ABSTRACT]{\justifying
SSuperconducting fluctuations (SF), arising from Cooper-pair-driven interactions and electron-electron interactions within the transition regime, can dominate the magnetoconductance (MC) of superconducting thin films near their transition temperature. In some systems, this fluctuation-dominated regime is preceded by a narrow window in which quantum interference effects are manifested as weak antilocalization (WAL) or weak localization (WL). Here, we report magnetotransport measurements on thin films of the high-temperature superconductor Bi$_2$Sr$_2$CaCu$_2$O$_{8+\delta}$ (BSCCO) that establish SF as the dominant mechanism governing MC across the transition regime. In the narrow temperature window well above $T_c$ where SF does not exist, a cusp in MC can be attributed to WAL-like signature well described by the Hikami--Larkin--Nagaoka (HLN) framework. Closer to the transition ($T_{\mathrm{BKT}}$), we observe signatures consistent with Berezinskii--Kosterlitz--Thouless (BKT) correlations, underscoring that the dominant physics is set by collective, pairing-driven behavior rather than single-particle quantum interference. Measurements on non-superconducting control samples show no WAL-like features, indicating that the WAL over narrow temperature window, emerges only in conjunction with superconducting fluctuations. Hall measurements further reveal a concomitant sign reversal of the dominant carrier response, pointing to a substantial reorganization of transverse transport coincident with the onset of superconductivity. Together, these observations establish that magnetotransport in BSCCO thin films near the superconducting transition is primarily governed by superconducting-fluctuation and vortex-dynamics physics, with quantum interference playing at most a subordinate, non-universal role confined to a narrow precursor regime. Our results motivate theoretical treatments that prioritize the interplay between superconducting correlations and collective pairing excitations over quasiparticle interference in describing quantum-coherent transport in cuprate superconductors.}

\maketitle

% ===============================================================
\section{Introduction}
% ===============================================================

\noindent\justifying
Quantum corrections to the normal-state conductivity provide a sensitive probe of the underlying electronic correlations in low-dimensional and disordered systems~\cite{Bergmann1984,Giannouri1997,ALTSHULER1985,Lee1985,Yadav2024,Wang2022}. In materials proximate to a superconducting instability these corrections are frequently dominated by fluctuation effects rather than by weak localization phenomena alone. Above the superconducting transition temperature, $T_c$, thermodynamic fluctuations of the superconducting order parameter give rise to transient Cooper pairs, which provide an additional conduction channel known as superconducting fluctuations (SF). This results in excess conductivity comprising contributions from the Aslamazov--Larkin (AL) and Maki--Thompson (MT) correction terms~\cite{ALTSHULER1985}. These fluctuation-induced corrections are strongly temperature- and field-dependent, and in systems with strong spin--orbit coupling or specific disorder configurations they can coexist with, or even dominate, quantum interference corrections~\cite{ALTSHULER1985,Yadav2024,Wang_2022}. While the Hikami--Larkin--Nagaoka (HLN) formalism~\cite{HLN_1980} is conventionally invoked to describe the low-field magnetoconductance (MC) cusp in such systems, a cusp-like feature of this form is not by itself sufficient to establish a WAL origin, since AL- and MT-type fluctuation corrections can produce a qualitatively similar MC signature in the vicinity of $T_c$.

\noindent
This ambiguity becomes particularly relevant at elevated temperatures, where genuine WAL is expected to be strongly suppressed: phase coherence lengths shrink rapidly once inelastic electron--phonon and electron--electron scattering set in. In a purely normal-state picture, WAL-like corrections should survive only over a narrow window of temperature and field except in materials with unusually strong spin--orbit coupling or unconventional band structures~\cite{Li2023,Salawu2022,KUMAR2021}.

In superconductors, the persistence of a cusp in MC well beyond this expected window is therefore more naturally understood as a signature of superconducting fluctuations that grow in strength as $T \to T_c$, rather than as an intrinsic weak anti-localization effect. This motivates a quantitative analysis in which the low-field MC is examined not solely within the HLN framework but as a possible crossover between a WAL-like quantum correction at temperatures well above $T_c$ and fluctuation-dominated (AL/MT) regime on approach to the superconducting transition~\cite{ALTSHULER1985}.

\noindent
High-temperature superconductors provide a fertile platform for exploring such unconventional quantum transport phenomena~\cite{Ayres2021,Pelc2019,Ren2008,Yurii2008}. Among them, Bi$_2$Sr$_2$CaCu$_2$O$_{8+\delta}$ (BSCCO) is a prototypical layered superconductor characterized by strong electronic correlations, pronounced anisotropy, and a complex Fermi surface that evolves with temperature and doping~\cite{Drozdov2018,Shen1991,Zareapour2012}. Transport in BSCCO thin films often exhibits quasi-two-dimensional behavior, making them well suited for investigating quantum interference phenomena in close proximity to SF~\cite{Wang2023,Kotzler1997}. Fundamentally, the superconducting transition in such systems is described by a Berezinskii--Kosterlitz--Thouless (BKT) mechanism, associated with the unbinding of vortex--antivortex pairs and characterized by a transition temperature $T_{\mathrm{BKT}}$~\cite{Wang2023,Kotzler1997,Ying1990,Mondal2011,Yong2013,Weitzel2023}.

Despite decades of intense investigation, however, a complete microscopic understanding of the normal-state and superconducting properties of the cuprates remains elusive. Whether the concepts developed for simple metals and BCS-like superconductors, retain their validity in a strongly correlated high-$T_c$ superconductors is itself an open and largely unanswered question. In particular, the pairing mechanism in cuprate superconductors remains unresolved, and the electronic and transport properties in the vicinity of $T_c$ are considerably more complex than in conventional, weakly correlated superconductors, reflecting the interplay of strong correlations, pseudogap physics, and competing orders~\cite{Ayres2021,Pelc2019,Drozdov2018}.

\noindent
A crossover from WAL/WL-like feature to a rapidly growing SFs contribution has been reported in ultra-thin conventional $s$-wave superconductors such as NbN~\cite{Weitzel2023}, TiN~\cite{Yadav2024} and Sn$_{1-x}$In$_x$Te~\cite{Wang_2022}. Fluctuation phenomena in cuprates are further complicated by strong electronic correlations, a $d$-wave pairing symmetry. SFs are indeed reported in BSCCO, however, the measurement methods are spectroscopic rather than via electronic transport~\cite{Yu_2021,Chen_2019}, our work aims to address this lacunae.

\noindent
In this work, we report a smooth crossover from WAL-like MC to an SF-dominated regime in large-area, sputtered thin films of Bi$_2$Sr$_2$CaCu$_2$O$_{8+\delta}$. We demonstrate that the appearance of WAL-like features is confined to a narrow temperature window immediately preceding the onset of the superconducting transition, where the low-field MC is well described by the HLN framework. As superconducting correlations strengthen, the MC deviates from HLN behavior and crosses over into the fluctuation-dominated regime. Measurements on non-superconducting control samples show no WAL-like features, indicating that these signatures arise specifically in conjunction with superconducting correlations. In addition, Hall measurements reveal a concomitant sign reversal of the dominant carrier response, suggesting a substantial modification of transverse transport coincident with the onset of superconductivity.

% ===============================================================
\section{Results and Discussion}
% ===============================================================

\begin{figure*}[!t]
\centering
\includegraphics[width=\textwidth]{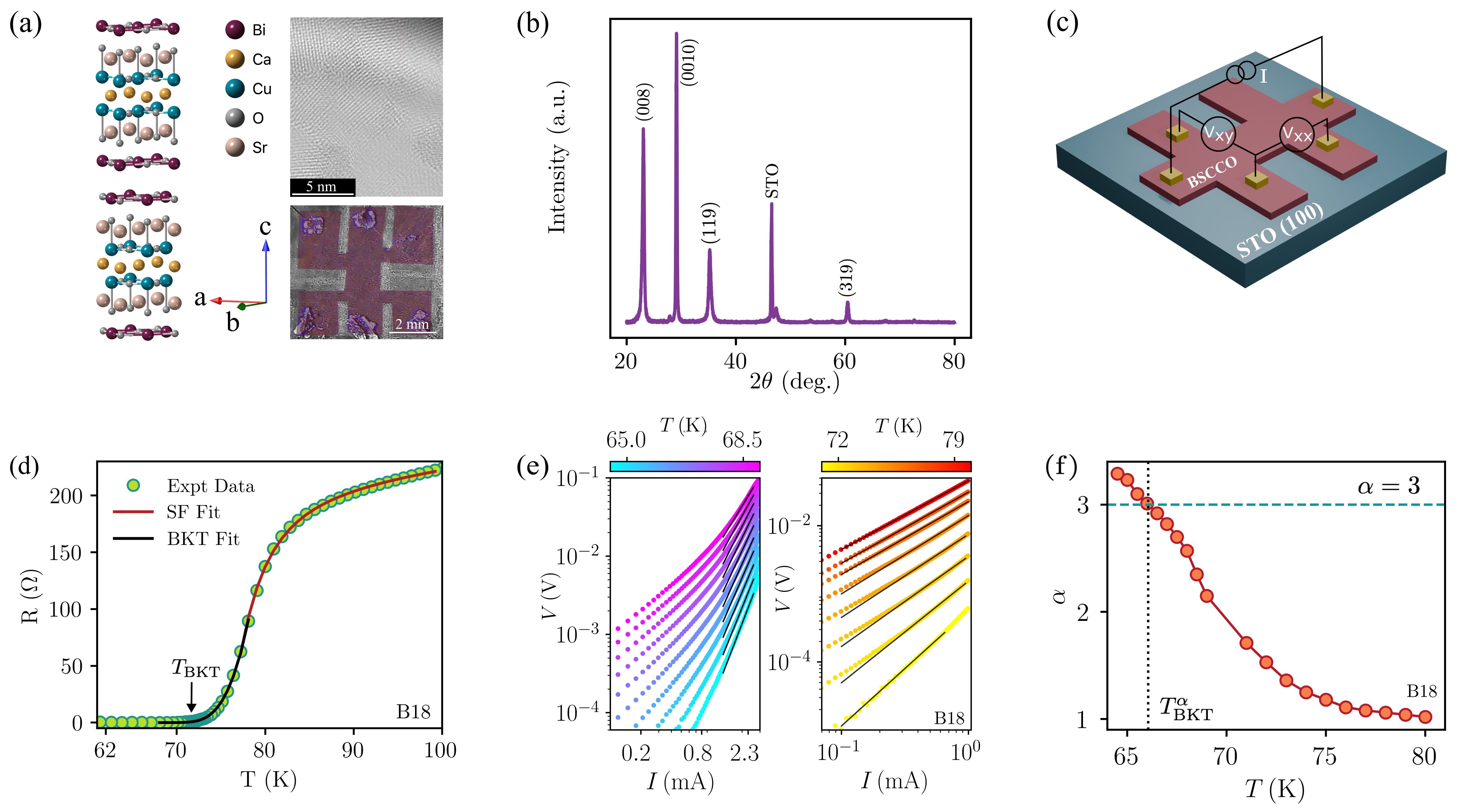}
\caption{Crystal structure and transport measurements of BSCCO thin film: (a) Crystal structure of the 2212 phase of BSCCO generated using \textit{CrystalMaker} software (left panel). Cross-sectional transmission electron microscopy (TEM) image of the BSCCO crystal, revealing atomic-scale lattice ordering (top right panel). False color optical microscope image of the Hall bar of BSCCO used for measurements (bottom right panel). (b) X-ray diffraction spectrum showing 2212 phase of BSCCO. (c) Schematic of the Hall bar measurement configuration. (d) Temperature-dependent resistance of the BSCCO thin film (thickness $t = 35~\mathrm{nm}$), with data fitted to SF model [eqn.~\eqref{eq:RT_fit}] and Halperin--Nelson (HN) model [eqn.~\eqref{eq:HN}] as indicated in the legend. (e) Current--voltage ($I$--$V$) characteristics measured at different temperatures, plotted on a log--log scale. The left panel shows low-temperature data ($65.5$--$69~\mathrm{K}$); the right panel shows high-temperature data ($71$--$80~\mathrm{K}$). Black solid lines are linear fits using $\log V = \alpha \log I$, corresponding to the power-law behavior $V \propto I^{\alpha}$. (f) Temperature dependence of the extracted exponent $\alpha$. The blue dashed line marks the universal BKT criterion $\alpha = 3$; its intersection with the $\alpha(T)$ curve defines the BKT transition temperature $T_{\mathrm{BKT}}^{\alpha}$. Results shown are for sample B18, as indicated in the figures.}
\label{intro}
\end{figure*}

We initiate our investigation of the BSCCO samples by characterizing their crystal structure. The unit cell of 2212 phase, as obtained via \textit{CrystalMaker} simulations is shown in fig.~\ref{intro}(a). Transmission electron microscopy (TEM) imaging reveals crystal lattices with incommensurate supermodulation predominantly oriented along the $(119)$ crystallographic axis. Cross-sectional TEM images demonstrate polycrystalline ordering of atomic sites [fig.~\ref{intro}(a)]. Our samples exhibit 2212 phase of BSCCO, as seen via X-ray diffraction [fig.~\ref{intro}(b)] and TEM microscopy over multiple regions (see SI).

We probe the electronic transport properties using the standard four-point measurement configuration [Fig.~\ref{intro}(c)] by measuring the resistance ($R$) as a function of temperature ($T$) [Fig.~\ref{intro}(d)]. The temperature range of $78$--$88~\mathrm{K}$ lies in the immediate vicinity above the superconducting transition, where thermally excited Cooper pairs are formed but long-range phase coherence has not yet been established. In this fluctuation regime, the conductivity acquires additional contributions from SFs: the Aslamazov--Larkin (AL)~\cite{Aslamasov1968,Redi1977}, Maki--Thompson (MT)~\cite{Larkin1980,Thompson1970,Maki1968}, and density-of-states (DOS) terms~\cite{Dorin1993,Ioffe1993}. The total conductance is
\begin{equation}
G(T)=G_{\mathrm{normal}}(T)+\Delta G_{\mathrm{AL}}(T)+\Delta G_{\mathrm{MT}}(T)+\Delta G_{\mathrm{DOS}}(T),
\label{eq:total_cond}
\end{equation}
where the normal-state conductance is
\begin{equation}
G_{\mathrm{normal}}(T)=\frac{1}{R_0+AT},
\label{eq:Gnormal}
\end{equation}
with $R_0$ and $A$ the fitting parameters for the residual resistance (zero-temperature intercept) and temperature coefficient of resistance, respectively.

The AL term describes direct transport of thermally generated fluctuating Cooper pairs,
\begin{equation}
\Delta G_{\mathrm{AL}}(T)=\frac{A_{\mathrm{AL}}}{\ln(T/T_c)},
\label{eq:AL}
\end{equation}
where $A_{\mathrm{AL}}$ is an effective fitting amplitude incorporating the theoretical prefactors, and $T_c$ is the mean-field transition temperature below which the order parameter acquires finite amplitude.

The MT term arises from coherent quasiparticle scattering mediated by fluctuating Cooper pairs,
\begin{equation}
\Delta G_{\mathrm{MT}}(T)=A_{\mathrm{MT}}\ln\!\left(\frac{k_{\mathrm{B}}T\tau_{\phi}}{\hbar}\right),
\label{eq:MT}
\end{equation}
where $\tau_{\phi}$ is the phase-coherence (dephasing) time and $A_{\mathrm{MT}}$ the corresponding amplitude.

The DOS term accounts for the reduced quasiparticle density of states due to transient Cooper-pair formation,
\begin{equation}
\Delta G_{\mathrm{DOS}}(T)=A_{\mathrm{DOS}}\ln\!\left[\frac{\left|\ln(T/T_c)\right|}{\left|\ln\!\left(\dfrac{k_{\mathrm{B}}T_c\tau}{\hbar}\right)\right|}\right],
\label{eq:DOS}
\end{equation}
where $\tau$ is the quasiparticle scattering time and $A_{\mathrm{DOS}}$ the corresponding amplitude. $A_{\mathrm{AL}}$, $A_{\mathrm{MT}}$, and $A_{\mathrm{DOS}}$ are phenomenological amplitudes absorbing the theoretical prefactors while accounting for material-specific properties and experimental uncertainties.

The experimental $R(T)$ data in Fig.~\ref{intro}(d) were fitted over $78$--$88~\mathrm{K}$ using the sum of all SF contributions,
\begin{equation}
R(T)=\frac{1}{G_{\mathrm{normal}}(T)+\Delta G_{\mathrm{AL}}(T)+\Delta G_{\mathrm{MT}}(T)+\Delta G_{\mathrm{DOS}}(T)},
\label{eq:RT_fit}
\end{equation}
with the red solid line showing the best fit. The normal-state conductance was obtained from a linear fit to the high-temperature resistance. The combined SF fit yields $T_c=76.19~\mathrm{K}$, with extracted scattering time $\tau=2.83\times10^{-15}~\mathrm{s}$ and phase-coherence time $\tau_{\phi}=2.06\times10^{-9}~\mathrm{s}$. The fitting parameters $\tau$ and $\tau_{\phi}$ were constrained to physically reasonable ranges reported for disordered conductors and superconductors, where the elastic scattering time is typically of the order of $10^{-15}$--$10^{-14}$~s and the phase-coherence time ranges from $\sim10^{-11}$ to $10^{-9}$~s~\cite{Truccato_2006,Lee1985,Altshuler1998,Skvortsov2004}. The extracted amplitudes $A_{\mathrm{AL}}$, $A_{\mathrm{MT}}$, and $A_{\mathrm{DOS}}$ are consistent with previously reported analyses of SF conductivity in disordered superconductors, where the AL, MT, and DOS contributions are found to be of comparable magnitude~\cite{Larkinbook,Aslamasov1968,Maki1989}. The good agreement between experiment and the SF fit indicates that transport in the range $78$--$100~\mathrm{K}$ is governed primarily by SF.

Closer to the zero-resistance state, below $78~\mathrm{K}$, the superconducting amplitude is already well developed, while phase coherence is progressively established. Transport is then controlled by phase fluctuations associated with the proliferation and binding of vortex--antivortex pairs. The resistance in this regime is therefore dominated by vortex dynamics rather than by amplitude fluctuations of the order parameter. The Halperin--Nelson expression~\cite{Halperin1979}, derived for the BKT transition, explicitly captures the exponential suppression of resistance due to vortex--antivortex binding and is thus appropriate for modeling the $R(T)$ behavior close to the zero-resistance state. The Halperin--Nelson phase fluctuation expression is given by:
\begin{equation}
R = A \exp\left(\frac{-b}{\sqrt{T/T_{\mathrm{BKT}} - 1}}\right),
\label{eq:HN}
\end{equation}
where $T_{\mathrm{BKT}}$ is the BKT transition temperature and $A, b>0$ are fitting parameters. Applying this model over $\approx 68$--$78~\mathrm{K}$ yields $T_{\mathrm{BKT}} = 66.85~\mathrm{K}$. An excellent fit to the model implies that the transition is because of vortex-antivortex pairing and unbinding~\cite{Kosterlitz1973}; which is consistent with layered VdW nature of the material.

Near $T_{\mathrm{BKT}}$, the current--voltage ($I$--$V$) characteristics provide critical insights into the superconducting-to-normal-metal transition mechanism~\cite{Venditti2019,Weitzel2023}. The BKT transition is characterized by a power-law relation between voltage and current, $V \propto I^{\alpha}$, where $\alpha$ is the power-law exponent. The universal BKT transition is identified by the criterion $\alpha = 3$. The transport measurements performed on the BSCCO thin films exhibit this characteristic power-law scaling in the $I$--$V$ curves, consistent with the theoretical predictions of the BKT transition [Fig.~\ref{intro}(e)]. In the high-current regime, Ohmic dissipation, dominated by Joule heating, results in a linear $I$--$V$ response with finite resistance [right panel of Fig.~\ref{intro}(e)], corresponding to $\alpha \approx 1$ for $T>77~\mathrm{K}$~\cite{Weitzel2023}. Conversely, in the superconducting state, the voltage approaches zero due to the formation of bound vortex--antivortex pairs, which suppress free vortex motion and enable dissipationless current transport~\cite{Kosterlitz1973}. Near the superconducting phase boundary, the $I$--$V$ characteristics become strongly nonlinear and are well described by the BKT power-law scaling, as illustrated in the left panel of Fig.~\ref{intro}(e).

The temperature-dependent power-law exponent, $\alpha$, was extracted from linear fits to the $\log(V)$ versus $\log(I)$ curves [Fig.~\ref{intro}(f)]. The intersection of the experimentally determined $\alpha(T)$ curve with the universal line $\alpha = 3$ yields a BKT transition temperature of $T_{\mathrm{BKT}}^{\alpha} \approx 66.06~\mathrm{K}$. This value is in close agreement with $T_{\mathrm{BKT}} = 66.85~\mathrm{K}$ obtained independently from the resistive transition measurements, further confirming the occurrence of a BKT transition in the BSCCO thin films. This minor temperature discrepancy ($\approx 0.8$~K) between the two experimental approaches originates from fundamental differences in the vortex excitation mechanisms inherent to each measurement protocol: (i) $I$--$V$ characterization employs current ramping, inducing vortex penetration from sample peripheries at the critical current threshold; (ii) $R$ versus $T$ measurements utilize thermal activation, promoting homogeneous vortex nucleation throughout the sample volume. These distinct vortex generation pathways result in different effective transition temperatures, reflecting the sensitivity of $T_{\mathrm{BKT}}$ determination to the specific experimental methodology employed for probing the BKT phase transition.

\begin{figure*}[!t]
\centering
\includegraphics[width=\textwidth]{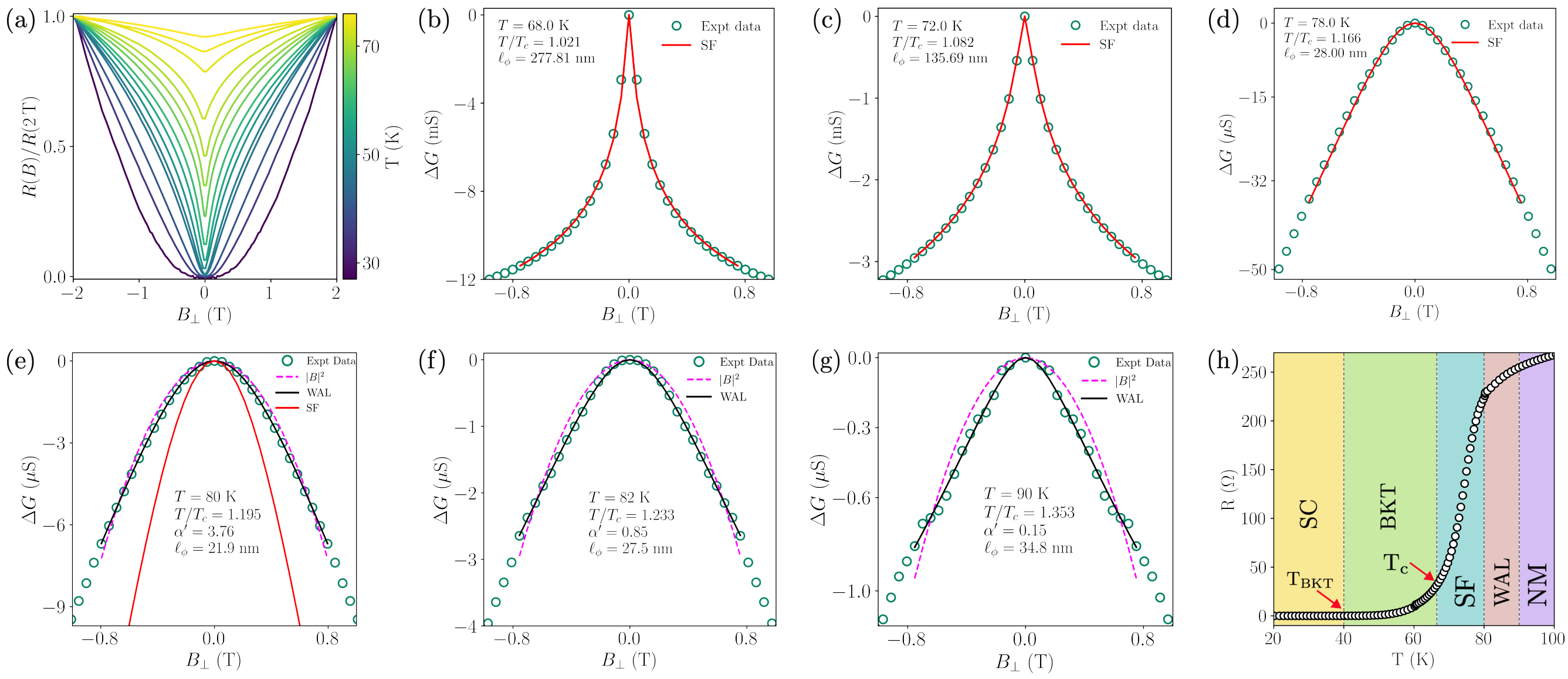}
\caption{(a) Normalized resistance of the sample as a function of out-of-plane magnetic field from $-2$~T to $+2$~T at different temperatures, as indicated by the color bar. (b--d) Magnetoconductance ($\Delta G$) versus out-of-plane magnetic field at 68~K, 72~K, and 78~K, respectively. The red solid lines represent fits to the SF model. (e) Magnetoconductance ($\Delta G$) versus out-of-plane magnetic field at 80~K, showing individual fits to the SF model (red solid line), a quadratic dependence ($-B^2$; magenta dotted line), and the HLN model (black solid line). (f--g) Magnetoconductance ($\Delta G$) as a function of the out-of-plane magnetic field at 82~K and 90~K, respectively, showing fits to the HLN model (black solid lines) and a quadratic dependence ($-B^2$; magenta dotted lines). (h) Resistance as a function of temperature, highlighting the different transport regimes identified from the analysis. These results are of sample B08D. The $R$ vs $T$ of this sample is shown in the Fig.~S3, SI.}
\label{wal}
\end{figure*}

Figure~\ref{wal}(a) shows the normalized magnetoresistance, $R(B)$, of the BSCCO thin film measured with the magnetic field applied perpendicular to the film plane over a wide range of temperatures. At temperatures well below the superconducting transition temperature ($T_c \approx 66.5~\mathrm{K}$), the sample exhibits zero resistance up to the lower critical field ($H_{c1}$). Beyond $H_{c1}$, the resistance increases gradually with the magnetic field, approaching the normal-state resistance.

A pronounced low-field cusp in the magnetoconductance (MC) is observed across a broad temperature range, spanning the superconducting transition region and persisting to temperatures well above $T_c$. Similar low-field MC cusps have been attributed to several mechanisms, including superconducting fluctuations (SF)~\cite{Yadav2024,Giannouri1997}, vortex dynamics, Fermi-surface reconstruction~\cite{Koshelev2024}, and quantum interference effects such as weak anti-localization (WAL)~\cite{Hikami1980}. In the present study, the absence of any clear charge density wave-related features in the $R(T)$ characteristics suggests that Fermi-surface reconstruction is unlikely to be the primary origin of the observed cusp, although it cannot be completely ruled out based on transport measurements. Therefore, to identify the dominant mechanism responsible for the low-field MC cusp, the experimental data were systematically analyzed using various theoretical models describing SF and quantum interference effects.

The applicability of these models depends on the dimensionality of the system. Since the phase coherence length extracted from the MC analysis at 68~K ($\ell_{\phi} \approx 277$~nm) is much larger than the film thickness, and the $R(T)$ analysis together with the non-linear $I$--$V$ characteristics exhibit BKT behavior, the BSCCO thin film can be treated as a quasi two-dimensional superconductor. This is further supported by an electron scattering time of $\tau \approx 1.8\times10^{-15}$~s and a dephasing time of $\tau_{\phi}\approx2.4\times10^{-10}$~s extracted from the $R(T)$ analysis, satisfying the condition $\tau_{\phi}\gg\tau$~\cite{Lee1985,Aleiner1999,Brunthaler2001} required for quantum interference effects. Therefore, two-dimensional theories of SF and WAL are appropriate for describing the present MC data.

In two-dimensional superconductors, SF persist above $T_c$ due to the formation of transient Cooper pairs, giving rise to a fluctuation-induced excess conductivity. This excess conductivity is primarily described by two contributions: the Aslamazov--Larkin (AL) term~\cite{Aslamasov1968,Redi1977}, which originates from the direct formation of transient Cooper pairs and dominates very close to $T_c$ [$(T-T_c)/T_c\ll1$], and the Maki--Thompson (MT) term~\cite{Larkin1980,Thompson1970,Maki1968}, which arises from the coherent scattering of quasiparticles by fluctuating Cooper pairs before phase coherence is lost. Since both the AL and MT contributions are suppressed by an applied magnetic field, the fluctuation-induced conductivity decreases with increasing field, producing a characteristic low-field MC cusp.

A similar low-field cusp can also originate from WAL, a quantum interference effect arising from the destructive interference of time-reversed electron trajectories in the presence of strong spin--orbit coupling. An applied magnetic field suppresses this interference correction and consequently reduces the conductivity. Since SF and WAL give rise to qualitatively similar low-field MC signatures, distinguishing between these mechanisms based solely on the experimental MC curves is challenging.

Based on these considerations, the MC data were analyzed using theoretical models incorporating three quantum corrections to the conductivity: the AL contribution, the MT contribution, and WAL. Their relative contributions across different temperature regimes are discussed in the following sections.

The MC can be expressed as the sum of the SF and WAL contributions,
\begin{equation}
\Delta G_{xx}(B)=G_{xx}(B)-G_{xx}(0)=\Delta G_{\mathrm{SF}}(B)+\Delta G_{\mathrm{WAL}}(B),
\label{eq:MC_total}
\end{equation}
where $\Delta G_{\mathrm{SF}}$ represents the conductivity correction arising from SF, while $\Delta G_{\mathrm{WAL}}$ corresponds to the quantum interference correction associated with WAL.

The WAL contribution is described by the HLN model~\cite{Hikami1980},
\begin{equation}
\Delta G_{\mathrm{WAL}}(B)
=
-\frac{\alpha^{\prime} e^{2}}{\pi h}
\left[
\ln\left(\frac{h}{8\pi B e \ell_{\phi}^{2}}\right)
-
\Psi\left(
\frac{1}{2}
+
\frac{h}{8\pi B e \ell_{\phi}^{2}}
\right)
\right],
\label{eq:HLN}
\end{equation}
where $\ell_{\phi}$ is the phase coherence length, $\Psi$ is the digamma function, and $\alpha^{\prime}$ is a dimensionless prefactor that characterizes the number of coherent transport channels.

The $\Delta G(B)$ curves in the temperature range of 90--82~K are well described by the HLN model, as shown in Figs.~\ref{wal}(g) and \ref{wal}(f), yielding excellent agreement with the experimental data. The genesis of WAL may be attributed to spin orbit coupling arising from heavy Bi atoms and the broken inversion symmetry at the film surface and the film/SrTiO$_3$ interface. The extracted prefactor $\alpha^{\prime}$ remains below unity ($0.85$ at 82~K, $0.15$ at 90~K) throughout this temperature range [Fig.~\ref{wal}(e)], consistent with transport dominated by weak antilocalization (WAL). In this regime, the WAL-induced MC correction is typically of the order of $e^{2}/h$ (approximately a few tens of $\mu\mathrm{S}$). However, below 80~K, the HLN model fails to adequately describe the experimental MC. The fitted prefactor $\alpha^{\prime}$ exceeds unity, reaching unphysically large values, along with the drastic increase of the magnitude of $\Delta G$ from the $\mu\mathrm{S}$ scale to the $\mathrm{mS}$ scale. The increase in $\alpha^{\prime}$ suggests the emergence of additional contributions to the MC. While it may indicate multiple quantum-coherent transport channels~\cite{Shyaga2025} arising from the layered crystal structure and structural inhomogeneity of the sputtered BSCCO thin films, it also reflects the increasing contribution of SF as the temperature approaches $T_c$. Consequently, the enhancement of $\alpha^{\prime}$ indicates that both WAL and SF can be contributing simultaneously to the MC in this temperature regime. Non-superconducting control films show no such cusp, so we refer to this feature as WAL-like and to HLN as the functional form that describes it [see SI].

We have further compared a quadratic field dependence, $\Delta G(B)\propto-aB^2$, with the magnetoconductance data in the 82--90~K range [Fig.~\ref{wal}(e--g)], using the same magnetic-field interval ($-0.8$ to $+0.8$~T) as in the HLN analysis. The quadratic model fails to reproduce the measured field dependence adequately, whereas the Hikami--Larkin--Nagaoka (HLN) expression provides a better description. This suggests that the observed magnetoconductance cannot be adequately described by a simple quadratic orbital contribution alone.

The SF contribution, $\Delta G_{\mathrm{SF}}$, consists of two main components: the AL and MT contributions~\cite{Bergmann1984,Giannouri1997,Harris2018,Yadav2024,Wang2022}. The AL contribution to the MC is given by
\begin{equation}
\begin{split}
\Delta G^{\mathrm{AL}}(B,T)
&=
G_{00}
\frac{\pi^{2}}
{8\ln\left(\dfrac{T}{T_c}\right)}
\Bigg\{
8\left(\frac{B_{\mathrm{SF}}}{B}\right)^2 \\
&\quad\times
\left[
\psi\left(\frac{1}{2}+\frac{B_{\mathrm{SF}}}{B}\right)
-
\psi\left(1+\frac{B_{\mathrm{SF}}}{B}\right)
+
\frac{B}{2B_{\mathrm{SF}}}
\right]
-1
\Bigg\},
\end{split}
\label{eq:AL_MC}
\end{equation}
where $G_{00}=e^{2}/(2\pi^{2}\hbar)$ is the quantum conductance, $\psi$ is the digamma function, and $B_{\mathrm{SF}}$ is the characteristic magnetic field associated with SF. The characteristic field is related to the Ginzburg--Landau relaxation time, $\tau_{\mathrm{GL}}$, through
\begin{equation}
B_{\mathrm{SF}}
=
\frac{\hbar}{4eD\tau_{\mathrm{GL}}},
\label{eq:BSF}
\end{equation}
where
\begin{equation}
\tau_{\mathrm{GL}}
=
\frac{\pi\hbar}
{8k_{\mathrm{B}}T
\ln\left(T/T_c\right)},
\label{eq:tauGL}
\end{equation}
and the diffusion constant is given by
\begin{equation}
D
=
\frac{\pi}{2\gamma}
\frac{k_{\mathrm{B}}T_c}
{eB_{c2}(0)},
\qquad
\gamma=1.78,
\label{eq:8}
\end{equation}
where $B_{c2}(0)$ is the upper critical field at zero temperature.

The MT contribution to the MC is expressed as
\begin{equation}
\Delta G^{\mathrm{MT}}(B,T)
=
-G_{1}
\frac{e^{2}}
{2\pi^{2}\hbar}
\left[
\psi\left(\frac{1}{2}+\frac{B_{\phi}}{B}\right)
-
\psi\left(\frac{1}{2}+\frac{B_{\mathrm{SF}}}{B}\right)
+
\ln\left(\frac{B_{\mathrm{SF}}}{B_{\phi}}\right)
\right],
\label{eq:MT_MC}
\end{equation}
where
\begin{equation}
G_{1}
=
C\,\beta_{\mathrm{LdS}}(T/T_c,\delta),
\label{eq:G1}
\end{equation}
with
\begin{equation}
\beta_{\mathrm{LdS}}(T/T_c,\delta)
=
\frac{\pi^{2}}
{4(\epsilon-\delta)},
\qquad
\epsilon
=
\ln\left(\frac{T}{T_c}\right),
\qquad
\delta
=
\frac{\pi\hbar}
{8k_{\mathrm{B}}T\,\tau_{\phi}}.
\label{eq:beta}
\end{equation}
Here, $\beta_{\mathrm{LdS}}$ represents the Larkin electron--electron interaction parameter in the limit $\epsilon\ll1$, $\delta$ is the pair-breaking parameter determined by the dephasing time $\tau_{\phi}$~\cite{Giannouri1997,Harris2018}, and $C$ is an additional fitting coefficient introduced to improve the agreement between the theoretical model and the experimental data~\cite{Yadav2024}. Consequently, the effective strength of the MT contribution is characterized by the parameter $G_{1}$.

The SF contribution to the MC is given by
\begin{equation}
\Delta G_{\mathrm{SF}}(B)
=
\Delta G_{\mathrm{AL}}(B)
+
\Delta G_{\mathrm{MT}}(B),
\label{eq:SF}
\end{equation}
This expression was used to fit the MC data in the temperature range of 68--80~K, as shown in Figs.~\ref{wal}(b)--\ref{wal}(e). The fitting parameters included the superconducting transition temperature ($T_c$), the MT fitting parameter ($C$), the dephasing time ($\tau_{\phi}$), and the diffusion constant ($D$). Since $D$ is related to the zero-temperature upper critical field, $H_{c2}(0)$ [Eq.~\eqref{eq:8}], which for high-$T_c$ cuprates such as BSCCO is typically reported to lie in the range of 50 and 150~T~\cite{Wang2003,Zavaritsky2001}, the corresponding $D$ was constrained to the range of $10^{-4}$--$10^{-6}$~m$^2$/s during the fitting procedure.

The superconducting transition temperature obtained from the $R(T)$ measurements is approximately $T_c \approx 66.5$~K. The extracted dephasing time, $\tau_{\phi}$, increases from approximately $4\times10^{-11}$~s to $4\times10^{-10}$~s as the temperature approaches $T_c$. Using the extracted values of $D$ and $\tau_{\phi}$, the phase coherence length was calculated from~\cite{Hikami1980,Lee1985}
\begin{equation}
\ell_{\phi}=\sqrt{D\tau_{\phi}}.
\label{eq:lphi}
\end{equation}
$\ell_{\phi}$ increases with decreasing temperature, reaching approximately 278~nm at 68~K, indicating the development of long-range phase coherence as the system approaches $T_c$.

Overall, the experimental MC in the temperature range of 68~K--78~K is well described by the SF model. However, it fails to adequately describe the MC data at 80~K [Fig.~\ref{wal}(e)]. At the same temperature, the prefactor $\alpha$ extracted from the HLN fit exceeds unity, indicating that the MC cannot be explained by the WAL or SF model alone. These observations suggest a narrow crossover region around 80~K, where SF and WAL coexist and contribute simultaneously to the observed MC.

Figure~\ref{wal}(h) summarizes the different transport regimes identified from the combined analysis of the temperature-dependent resistance and MC measurements. At temperatures well above the superconducting transition ($T \gtrsim 90$~K), the sample exhibits normal metallic (NM) behavior, where the $R(T)$ curve shows an approximately linear temperature dependence (refer to SI).

Upon cooling below $\sim90$~K, the transport enters a WAL regime spanning approximately 80~K to 90~K. In this temperature range, the MC is well described by the HLN model, indicating that quantum interference effects dominate the low-field transport.

As the temperature is further reduced below $\sim80$~K, SF become significant, and the MC is well described by the SF model, indicating that fluctuation-induced Cooper pairs provide the dominant contribution to the conductivity. Close to $T_c \approx 66.5$~K, a finite density of Cooper pairs develops, giving rise to a finite amplitude of the superconducting order parameter even though long-range phase coherence has not yet been established.

The temperature range between $T_{\mathrm{BKT}}$ and $T_c$ corresponds to the BKT regime. The BKT analysis of the $R(T)$ data yields a transition temperature of $T_{\mathrm{BKT}} \approx 39.9$~K (see SI). In this regime ($\sim39.9$~K--66.5~K), a finite superconducting order-parameter amplitude has already developed, while strong phase fluctuations prevent the establishment of long-range phase coherence. Below $T_{\mathrm{BKT}}$, vortex--antivortex pairs become bound, suppressing phase fluctuations and establishing long-range phase coherence, thereby resulting in a fully developed superconducting state with zero resistance.

Although Fig.~\ref{wal}(h) divides the transport behavior into distinct temperature regimes for clarity, the transitions between these regimes are expected to be gradual rather than abrupt. Consequently, narrow crossover regions may exist in which adjacent quantum transport mechanisms coexist.

\begin{figure}[!tbh]
\centering
\includegraphics[width=\columnwidth]{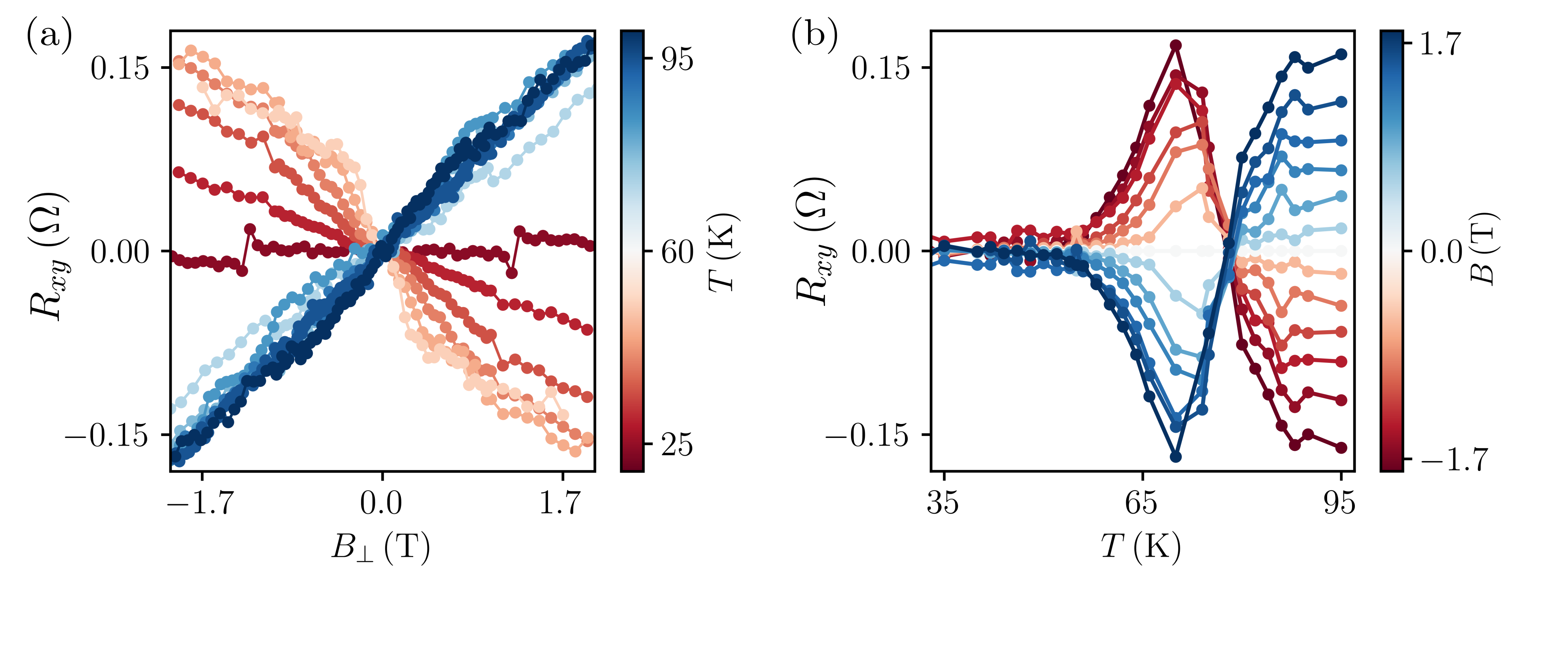}
\caption{(a) Hall resistance ($R_{xy}$) versus out of plane magnetic field ($B_\perp$) at different temperature as indicated by the color bar. (b) Hall resistance ($R_{xy}$) versus temperature at different out of plane magnetic field ($B_\perp$) as indicated by the color bar. These results are of sample B08D.}
\label{Hall}
\end{figure}

\noindent
We next examine the temperature dependence of the Hall resistance in the BSCCO thin-film samples [Fig.~\ref{Hall}(a)]. In conventional metals and semiconductors dominated by a single carrier type, the Hall carrier concentration is expected to be largely temperature independent, reflecting a stable carrier density and an essentially unchanged Fermi surface topology~\cite{Ashcroft1976}. In marked contrast, the superconducting BSCCO films exhibit a strong temperature dependence in the slope of the Hall resistance. This behavior is most clearly evidenced by the temperature at which $R_{xy}$ undergoes a sign reversal [Fig.~\ref{Hall}(b)]. Notably, in the vicinity of the superconducting transition temperature, the Hall carrier concentration changes sign from hole-like to electron-like. At sufficiently low temperatures, the Hall resistance vanishes concomitant with the zero-resistance state. Such a Hall sign reversal near the superconducting transition indicates a breakdown of simple single-band transport and points to the participation of additional transport channels or collective excitations~\cite{Sterpetti2017}. Hall sign reversal has previously been reported in cuprate superconductors and has been attributed to vortex dynamics governed by the Lorentz force and pinning effects~\cite{Zhao2019}. Moreover, superconducting fluctuations and vortex-like excitations are known to strongly modify transverse transport coefficients above and near $T_c$, giving rise to anomalous Hall responses~\cite{Pelc2020}. This behavior may reflect a competition between normal-state quasiparticle transport and fluctuation-induced channels associated with Cooper pairing and vortex motion. Together, the correlated appearance of WAL-like MC and Hall sign reversal indicates a profound restructuring of the transport landscape in the vicinity of the superconducting transition, calling for theoretical descriptions that explicitly incorporate the interplay between quasiparticles, superconducting fluctuations, and vortex dynamics in cuprate thin films.

% ===============================================================
\section{Conclusion}
% ===============================================================

\noindent
In summary, our magnetotransport study of large-area BSCCO thin films demonstrates that the approach to superconductivity in this system is governed primarily by SF rather than by only quantum interference. While a WAL-like correction to the MC is observed, it is confined to a narrow temperature window immediately preceding the SF, is well captured by the HLN framework only within that window, and breaks down rapidly as superconducting correlations strengthen. Furthermore, the results demonstrate a quasi two-dimensional transport as observed via the BKT transition in our thin films. This motivates future theoretical treatments of cuprate magnetotransport that prioritize the interplay between SF and interference corrections, rather than treating the two on equal footing. In addition, the large-area films synthesized and used in the experiments here are readily suitable for nano-fabrication, while the relatively straightforward growth technique offers a practical route toward realizing functional cryogenic devices~\cite{InglaAynes2025,Castellani2025,Ghosh2024} at high temperatures.

% ===============================================================
\section{Experimental Section}
% ===============================================================

Thin films of Bi$_2$Sr$_2$CaCu$_2$O$_{8+\delta}$ (BSCCO) with thicknesses ranging from 35~nm to 100~nm were deposited using radio-frequency (RF) sputtering. A base pressure of $\approx 10^{-8}$~mbar was achieved prior to deposition. Films were grown at a working pressure of $\sim 10^{-3}$~mbar with argon introduced at a flow rate of 40~sccm. The RF power was maintained at 100~W, corresponding to a deposition rate of approximately 8--10~nm~min$^{-1}$. All films were deposited on (100)-oriented SrTiO$_3$ substrates of size $5~\mathrm{mm} \times 5~\mathrm{mm}$ at room temperature. Following deposition, the films were annealed in an oxygen atmosphere at $850\,^{\circ}\mathrm{C}$ for 1~h to promote crystallization and oxygenation.

\subsection{Methodology}
Film thicknesses were determined using atomic force microscopy (AFM) (see Supporting Information for details). The samples were subsequently patterned into Hall bar geometries by mechanical scribing [lower panel of Fig.~\ref{intro}(a)], a procedure chosen to preserve sample integrity by avoiding chemical processing associated with conventional lithography. Electrical transport measurements were performed using a Keithley 6221 current source and 2182 nanovoltmeter in delta-mode configuration. Low-temperature measurements were carried out in a cryogen-free measurement system equipped with a superconducting magnet for applying out-of-plane magnetic fields. Anti-symmetrization of the Hall signal corresponds to $R_{xy} = \frac{R_{xy}(+B) - R_{xy}(-B)}{2}$. Therefore, Hall resistance was measured in both $+B$ and $-B$, so that any spurious even-in-field is eliminated.

% ---------------------------------------------------------------
\medskip
\noindent\textbf{Supporting Information}\par
\noindent\justifying
Additional supporting information can be found online in the Supporting Information section.
SupportingFile: SuppMat.docx.

\vspace{0.5cm}

% Acknowledgements
\medskip
\noindent\textbf{Acknowledgements}\par
\noindent\justifying
DS thanks IISc start-up grant, Ministry of Electronics and Information Technology (MeiTy), Govt.\ of India, Indian Space Research Organization (ISRO), Govt.\ of India for funding. Authors are grateful for funding from INOX Airproducts and INOXCVA for funding via CSR grants. DS thanks support from Infosys Foundation, Bangalore. PB thanks Anusandhan National Research Foundation, National Postdoctoral fellowship (PDF/2023/000444) for financial support. Authors are grateful to micro and nano characterization facility, CeNSE, IISc and national nano-fabrication facility, CeNSE, IISc both supported by Government of India, for facilities usage.

\bibliography{references} % .bib file name (without extension)

% ---------------------------------------------------------------
% Supporting Information (appended as-is from SI_Revised.pdf)
% ---------------------------------------------------------------
\clearpage
\onecolumn
\includepdf[pages=-,pagecommand={\thispagestyle{empty}}]{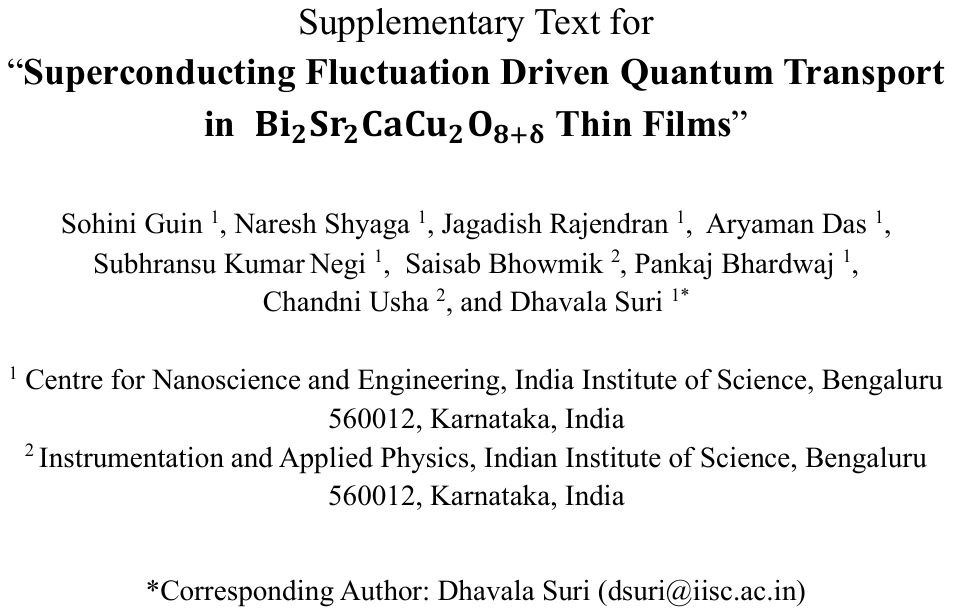}

\end{document}